\documentclass[a4paper]{article}

\usepackage{amsfonts}
\usepackage{amsbsy}
\usepackage{amssymb}
\usepackage{amsmath}
\usepackage{amsmath,amscd}
\usepackage{amsthm}
\usepackage{fancyhdr}
\usepackage{color}

\theoremstyle{plain}
\newtheorem{theorem}{Theorem}[section]
\newtheorem{lemma}{Lemma}[section]
\newtheorem{corollary}{Corollary}[section]
\newtheorem{proposition}{Proposition}[section]
\theoremstyle{definition}
\newtheorem{definition}{Definition}[section]
\newtheorem{remark}{Remark}[section]
\newtheorem{example}{Example}[section]

\renewcommand{\qed}{\hfill $\Box$}
\renewcommand{\proof}{{\em Proof. }}
\begin{document}

\title{Poisson structures on almost complex Lie algebroids}
\author{Paul Popescu}
\date{}
\maketitle

\begin{abstract}
In this paper we extend the almost complex Poisson structures from almost
complex manifolds to almost complex Lie algebroids. Examples of such
structures are also given and the almost complex Poisson morphisms of almost
complex Lie algebroids are studied.
\end{abstract}

\begin{flushleft}
\strut \textbf{2010 Mathematics Subject Classification:} 32Q60, 17B66, 53D17,

\textbf{Key Words:} almost complex structure, Lie algebroid, Poisson
structure
\end{flushleft}

\section{Introduction and preliminaries}

\setcounter{equation}{0}

\subsection{Introduction}

The Poisson manifolds are the smooth manifolds equipped with a Poisson
bracket on their ring of smooth functions. The importance of Poisson
structures both in real and complex geometry is without question. Short time
after their introduction in the real case \cite{Li1, W0} the Poisson
structures were studied in the context of some different categories in
differential geometry as: (almost) complex analytic category \cite{B-Z,
C-F-I-U, I-L, Ida1, L-S-X, L-S-X2, Li2, Pa1, Pa2, P, W}, foliated category 
\cite{B-C-I, Va5, Va7, Va8} or Banach category \cite{Ca-Pe, Ida0, O,O-R}.

In \cite{Li2} Lichnerowicz {introduces} the general concept of complex
Poisson structure, as a $2$--vector $\pi ^{2,0}$ of bidegree $(2,0)$ on a
complex manifold $M$ such that 
\begin{equation*}
\lbrack \pi ^{2,0},\pi ^{2,0}]=0\,,\,\,[\pi ^{2,0},\overline{\pi ^{2,0}}]=0.
\end{equation*}%
Related to this tensor, a bracket $\{\cdot ,\cdot \}$ on the algebra of
complex differentiable functions is defined, and when $\pi ^{2,0}$ is
holomorphic, this bracket can be reduced to the algebra of holomorphic
functions on the complex manifold, see also \cite{P}. These structures play
an important role in mathematics and mathematical physics, as well as the
holomorphic Poisson structures, for example, in the study of complex
Hamiltonian systems, see for instance \cite{B-C-R, P}. The local structure
of holomorphic Poisson manifolds was given in \cite{P}, just in the real
case \cite{W0}. Other generalizations as almost complex Poisson structures
and $\partial $-symplectic and $\partial $-Poisson structures was studied in 
\cite{C-F-I-U} and \cite{Pa1}, respectively, and also certain significant
studies concerning holomorphic Poisson manifolds in relation with
corresponding notions from the real case are given in some recent papers 
\cite{L-S-X, S}.

The Lie algebroids \cite{Mack, Mack2} are generalizations of Lie algebras
and integrable distributions. {\ For some algebraic extensions see, for
example, \cite{Ar}.} In fact a Lie algebroid is an anchored vector bundle
with a Lie bracket on module of sections. The cotangent bundle of a Poisson
manifold has a natural structure of a Lie algebroid and between Poisson
structures and Lie algebroids are many other connections, as for instance
for every Lie algebroid structure on an anchored vector bundle there is a
specific linear Poisson structure on the corresponding dual vector bundle
and conversely, see \cite{Va3, Va2}.

The almost complex Lie algebroids over almost complex manifolds were
introduced in \cite{B-R} as a natural extension of the notion of an almost
complex manifold to that of an almost complex Lie algebroid. This
generalizes the definition of an almost complex Poisson manifold given in 
\cite{C-F-I-U}, where some examples are also given. In \cite{I-Po} some
problems concerning to geometry of almost complex Lie algebroids over smooth
manifolds are studied in relation with corresponding notions from the
geometry of almost complex manifolds.

Also, we notice that the Poisson structures on real Lie algebroids are
introduced and studied, see for instance \cite{A-G-M, C-M, Fe, G-U, Kos,
Ma2, PoL2}.

The aim of this paper is to continue the study of Poisson structures, giving
a natural generalization of almost complex Poisson structures from almost
complex manifolds \cite{C-F-I-U} to almost complex Lie algebroids.

The paper is organized as follows. In the preliminary section, we briefly
recall some basic facts about Lie algebroids as well as the
Schouten-Nijenhuis bracket and Poisson structures on Lie algebroids. In the
second section we briefly present the almost complex Lie algebroids over
smooth manifolds and also recall the Newlander-Nirenberg theorem and almost
complex morphisms for almost complex Lie algebroids. Next we introduce the
notion of almost complex Poisson structures on almost complex Lie algebroids
and we describe some examples of such structures. Also, the complex
Lichnerowicz-Poisson cohomology of almost complex Poisson Lie algebroids is
introduced. We consider morphisms of almost complex Lie algebroids
preserving the almost complex Poisson structures that induce almost complex
Poisson structures on Lie subalgebroids. Also, we characterize the morphisms
of almost complex Poisson Lie algebroids in terms of its Graph which is a $%
J_{E}$--coisotropic Lie subalgebroid of the associated almost complex
Poisson Lie algebroid given by direct product structure.

{\ Compatibility conditions between complex structures and Poisson
structures arise naturally in the real form of holomorphic Poisson manifolds
and Lie algebroids; see , for example, \cite{L-S-X}. These compatibility
conditions are particular cases of Poisson-Nijenhuis structures as studied,
for example in \cite{G-U, Kos}. In the last section we give examples of
integrable almost complex structures and almost complex Poisson structures
that are \textit{not compatible} and that arise naturally on Lie algebroids
on each sphere $S^{2n-1}$, $n\geq 1$. These examples motivate the study of
almost complex Poisson Lie algebroids studied in this paper, where the
almost complex and the Poisson structures are not submitted to compatibility
conditions.  }

The main methods used here are similarly and closely related to those used
in \cite{C-F-I-U} for the case of almost complex Poisson manifolds.

\subsection{Preliminaries on Lie algebroids}

\begin{definition}
We say that $p:E\rightarrow M$ is an \textit{anchored} vector bundle if
there exists a vector bundle morphism $\rho:E\rightarrow TM$. The morphism $%
\rho$ will be called the anchor map.
\end{definition}

\begin{definition}
Let $(E,p,M)$ and $(E^{\prime},p^\prime,M^\prime)$ be two anchored vector
bundles over the same base $M$ with the anchors $\rho:E\rightarrow TM$ and $%
\rho^\prime:E^\prime\rightarrow TM$. A morphism of anchored vector bundles
over $M$ or a \textit{$M$--morphism of anchored vector bundles} between $%
(E,\rho)$ and $(E^\prime,\rho^\prime)$ is a morphism of vector bundles $%
\varphi:(E,p,M)\rightarrow(E^\prime,p^\prime,M)$ such that $%
\rho^\prime\circ\varphi=\rho$.
\end{definition}

The anchored vector bundles over the same base $M$ form a category. The
objects are the pairs $(E,\rho_E)$ with $\rho_E$ the anchor of $E$ and a
morphism $\phi:(E,\rho_E)\rightarrow(F,\rho_F)$ is a vector bundle morphism $%
\phi:E\rightarrow F$ which verifies the condition $\rho_F\circ\phi=\rho_E$.

Let $p:E\rightarrow M$ be an anchored vector bundle with the anchor $\rho
:E\rightarrow TM$ and the induced morphism $\rho _{E}:\Gamma (E)\rightarrow 
\mathcal{X}(M)$. Assume there {is} a bracket $[\cdot ,\cdot ]_{E}$ on {$%
\Gamma (E)$} that provides a structure of real Lie algebra on $\Gamma (E)$.

\begin{definition}
The triple $(E,\rho _{E},[\cdot ,\cdot ]_{E})$ is called a \textit{Lie
algebroid} if

\begin{enumerate}
\item[i)] $\rho_E:(\Gamma(E,[\cdot,\cdot]_E)\rightarrow(\mathcal{X}%
(M),[\cdot,\cdot])$ is a Lie algebra homomorphism, i.e. 
\begin{equation*}
\rho_E([s_1,s_2]_E)=[\rho_E(s_1),\rho_E(s_2)];
\end{equation*}

\item[ii)] $[s_1,fs_2]_E=f[s_1,s_2]_E+\rho_E(s_1)(f)s_2$,
\end{enumerate}

for every $s_1,s_2\in\Gamma(E)$ and $f\in C^{\infty}(M)$.
\end{definition}

A Lie algebroid $(E,\rho_E,[\cdot,\cdot]_E)$ is said to be \textit{transitive%
}, if $\rho_E$ is surjective.

There exists a canonical cohomology theory associated to a Lie algebroid $%
(E,\rho_E,[\cdot,\cdot]_E)$ over a smooth manifold $M$. The space $%
C^{\infty}(M)$ is a $\Gamma(E)$-module relative to the representation 
\begin{equation*}
\Gamma(E)\times C^{\infty}(M)\rightarrow
C^{\infty}(M),\,\,(s,f)\mapsto\rho_E(s)f.
\end{equation*}
Following the well-known Chevalley-Eilenberg cohomology theory \cite{C-E},
we can introduce a cohomology complex associated to the Lie algebroid as
follows. A $p$-linear mapping 
\begin{equation*}
\omega:\Gamma(E)\times\ldots\times\Gamma(E)\rightarrow C^{\infty}(M)
\end{equation*}
is called a $C^{\infty}(M)$-valued $p$-cochain. Let $\mathcal{C}^p(E)$
denote the vector space of these cochains.

The operator $d_E:\mathcal{C}^p(E)\rightarrow \mathcal{C}^{p+1}(E)$ given by 
\begin{equation}
(d_E\omega)(s_0,\ldots,s_p)=\sum_{i=0}^p(-1)^i\rho_E(s_i)(\omega(s_0,\ldots,%
\widehat{s_i},\ldots,s_p))  \label{I11}
\end{equation}
\begin{equation*}
+\sum_{i<j=1}^p(-1)^{i+j}\omega([s_i,s_j]_E,s_0,\ldots,\widehat{s_i},\ldots,%
\widehat{s_j},\ldots,s_p),
\end{equation*}
for $\omega\in \mathcal{C}^p(E)$ and $s_0,\ldots,s_p\in\Gamma(E)$, defines a
coboundary since $d_E\circ d_E=0$. Hence, $(\mathcal{C}^p(E),d_E)$, $p\geq1$
is a differential complex and the corresponding cohomology spaces are called
the cohomology groups of $\Gamma(E)$ with coefficients in $C^{\infty}(M)$.

\begin{lemma}
If $\omega\in \mathcal{C}^p(E)$ is skew-symmetric and $C^{\infty}(M)$%
-linear, then $d_E\omega$ also is skew-symmetric.
\end{lemma}

From now on, the subspace of skew-symmetric and $C^{\infty}(M)$-linear
cochains of the space $\mathcal{C}^p(E)$ will be denoted by $\Omega^p(E)$
and its elements will be called \textit{$p$--forms} on $E$ (or $p$--section
forms on $E$).

As in \cite{Fe}, the \textit{Lie algebroid cohomology} $H^p(E)$ of $%
(E,\rho_E,[\cdot,\cdot]_E) $ is the cohomology of the subcomplex $%
(\Omega^p(E),d_E)$, $p\geq 1 $. It is a natural generalization of the Lie
algebra case, as studied in \cite{C-E}.

\begin{definition}
Let $(E,\rho_E,[\cdot,\cdot]_E)$ and $(E^\prime,\rho_{E^\prime},[\cdot,%
\cdot]_{E^\prime})$ be two Lie algebroids over $M$. A \textit{morphism of
Lie algebroids} over $M$, is a morphism $\varphi:(E,\rho_E)\rightarrow(E^%
\prime,\rho_{E^\prime})$ of anchored vector bundles with property that: 
\begin{equation}  \label{I1}
d_E\circ \varphi^*=\varphi^*\circ d_{E^{\prime}},
\end{equation}
where $\varphi^*:\Omega^p(E^{\prime})\rightarrow\Omega^p(E)$ is defined by 
\begin{equation*}
(\varphi^*\omega^{\prime})(s_1,\ldots,s_p)=\omega^{\prime}(\varphi(s_1),%
\ldots,\varphi(s_p)),\,\omega^{\prime}\in\Omega^p(E^{\prime}),\,s_1,%
\ldots,s_p\in\Gamma(E).
\end{equation*}
We also say that $\varphi$ is a \textit{$M$--morphism of Lie algebroids}.
\end{definition}

\begin{remark}
Alternatively, we say that $\varphi :(E,\rho _{E})\rightarrow (E^{\prime
},\rho _{E^{\prime }})$ is {an} \textit{$M$--morphism of Lie algebroids} if 
\begin{equation}
\varphi \left( \lbrack s_{1},s_{2}]_{E}\right) =\left[ \varphi
(s_{1}),\varphi (s_{2})\right] _{E^{\prime }}\,,\,\,\forall \,s_{1},s_{2}\in
\Gamma (E).  \label{I01}
\end{equation}
\end{remark}

The Lie algebroids over the same manifold $M$ and all $M$--morphisms of Lie
algebroids form a category, which {(via a forgetful structure)} is a
subcategory of the category of anchored vector bundles over $M$.

Let $(E,\rho_E,[\cdot,\cdot]_E)$ be a Lie algebroid over $M$. If we consider 
$(x^{i})$, $i=1,\ldots,n$ a local coordinates system on $M$ and $\{e_a\}$, $%
a=1,\ldots,m$ a local basis of sections on the bundle $E$, where $\dim M=n$
and $\mathrm{rank}\,E=m$, then $(x^{i},y^{a})$, $i=1,\ldots,n$, $%
a=1,\ldots,m $ are local coordinates on $E$. Also, for an element $y\in E$
such that $p(y)=x\in U\subset M$, we have $y=y^{a}e_a(p(y))$.

In a such local coordinates system, the anchor $\rho_E$ and the Lie bracket $%
[\cdot,\cdot]_E$ are expressed by the smooth functions $\rho^{i}_a$ and $%
C^a_{bc}$, namely 
\begin{equation}  \label{I2}
\rho_E(e_a)=\rho^{i}_a\frac{\partial}{\partial x^{i}}\,\,\mathrm{and}%
\,\,[e_a,e_b]_E=C^c_{ab}e_c\,,\,i=1,\ldots,n,\,a,b,c=1,\ldots,m.
\end{equation}
The functions $\rho^{i}_a\,,\,C^{a}_{bc}\in C^\infty(M)$ given by the above
relations are called the \textit{structure functions} of Lie algebroid $%
(E,\rho_E,[\cdot,\cdot]_E)$ in the given local coordinates system.

If $(E,\rho _{E},[\cdot ,\cdot ]_{E})$ is a Lie algebroid over $M$, $(x^{i})$%
, $i=1,\ldots ,n$ is a local coordinate system on $M$ and $\{e_{a}\}$, $%
a=1,\ldots ,m$ is a local basis of sections on $E$, then the structure
functions $\rho _{a}^{i}\,,\,C_{bc}^{a}\in C^{\infty }(M)$ of the Lie
algebroid $E$ verify the following relations: 
\begin{equation}
\rho _{a}^{j}\frac{\partial \rho _{b}^{i}}{\partial x^{j}}-\rho _{b}^{j}%
\frac{\partial \rho _{a}^{i}}{\partial x^{j}}=\rho
_{c}^{i}C_{ab}^{c}\,,\,C_{ab}^{c}=-C_{ba}^{c}\,,\,\sum_{cycl(a,b,c)}\left(
\rho _{a}^{i}\frac{\partial C_{bc}^{d}}{\partial x^{i}}+C_{ab}^{e}C_{ce}^{d}%
\right) =0.  \label{I3}
\end{equation}

The equations \eqref{I3} are called the \textit{structure equations} of Lie
algebroid $(E,\rho_E,[\cdot,\cdot]_E)$.

\subsection{Poisson structures on Lie algebroids}

As in the classical settings from the geometry of Poisson manifolds, \cite%
{Va1}, the Schouten-Nijenhuis bracket on a Lie algebroid $%
(E,\rho_E,[\cdot,\cdot]_E)$ is a $\mathbb{R}$--linear map $[\cdot,\cdot]_E:%
\mathcal{V}^\bullet(E)\rightarrow\mathcal{V}^{\bullet-1}(E)$ defined by 
\begin{eqnarray*}
[s_1\wedge\ldots\wedge s_p,t_1\wedge\ldots \wedge
t_q]_E&=&(-1)^{p+1}\sum_{i=1}^p\sum_{j=1}^q(-1)^{i+j}[s_i,t_j]_E\wedge
s_1\wedge\ldots\wedge\widehat{s_i}\wedge\ldots\wedge s_p \\
&&\wedge t_1\wedge\ldots\wedge\widehat{t_j}\wedge\ldots\wedge t_q,
\end{eqnarray*}
for every $s_i,t_j\in\Gamma(E)$, \cite{Kos}, where $\mathcal{V}%
^p(E)=\Gamma(\bigwedge^pE)$ is the set of $p$--sections of $E$.

Let us consider the bisection (i.e. contravariant, skew-symmetric, $2$%
--section) $\pi_E\in\Gamma(\bigwedge^2E)$ locally given by 
\begin{equation}  \label{I4}
\pi_E=\frac{1}{2}\pi^{ab}(x)e_a\wedge e_b.
\end{equation}

\begin{definition}
\cite{Kos} The bisection $\pi_E\in\Gamma(\bigwedge^2E)$ is called a \textit{%
Poisson bisection} on the Lie algebroid $(E,\rho_E,[\cdot,\cdot]_E)$ if $%
[\pi_E,\pi_E]_E=0$.
\end{definition}

If $(x^{i})$, $i=1,\ldots ,n$ is a local coordinate system on $M$ and $%
\{e_{a}\}$, $a=1,\ldots ,m$ is a local basis of sections on $E$, then the
condition $[\pi _{E},\pi _{E}]_{E}=0$ is locally expressed as 
\begin{equation}
\sum_{(a,e,d)}\left( \pi ^{ab}\rho _{b}^{i}\frac{\partial \pi ^{ed}}{%
\partial x^{i}}+\pi ^{ab}\pi ^{cd}C_{bc}^{e}\right) =0.  \label{I5}
\end{equation}%
If $\pi _{E}$ is a Poisson bisection on the Lie algebroid $(E,\rho
_{E},[\cdot ,\cdot ]_{E})$, then the pair $(E,\pi _{E})$ is called a \textit{%
Poisson Lie algebroid}. A corresponding Poisson bracket on $M$ is given by 
\begin{equation}
\{f,g\}=\pi _{E}\left( d_{E}f,d_{E}g\right) \,,\,\,f,g\in C^{\infty }(M).
\label{I6}
\end{equation}%
Also, we have the map $\pi _{E}^{\#}:\Omega ^{1}(E)\rightarrow \Gamma (E)$
defined by 
\begin{equation}
\pi _{E}^{\#}\omega =\imath _{\omega }\pi _{E}\,,\,\,\omega \in \Omega
^{1}(E).  \label{I7}
\end{equation}%
Let us consider now the bracket 
\begin{equation}
\lbrack \omega ,\theta ]_{\pi _{E}}=\mathcal{L}_{\pi _{E}^{\#}\omega }\theta
-\mathcal{L}_{\pi _{E}^{\#}\theta }\omega -d_{E}(\pi _{E}(\omega ,\theta )),
\label{I8}
\end{equation}%
where $\mathcal{L}$ denotes the Lie derivative in the Lie algebroids
calculus framework, \cite{M}, and $\omega ,\theta \in \Omega ^{1}(E)$. With
respect to this bracket and the usual Lie bracket on vector fields, the map $%
\widetilde{\rho }_{E}:\Gamma (E^{\ast })\rightarrow \mathcal{X}(M)$ defined
by 
\begin{equation}
\widetilde{\rho }_{E}=\rho _{E}\circ \pi _{E}^{\#},  \label{I9}
\end{equation}%
is a Lie algebra homomorphism, namely $\widetilde{\rho }_{E}\left( [\omega
,\theta ]_{\pi _{E}}\right) =[\widetilde{\rho }_{E}(\omega ),\widetilde{\rho 
}_{E}(\theta )]$. Also, the bracket $[\cdot ,\cdot ]_{\pi _{E}}$ satisfies
the Leibniz rule 
\begin{equation*}
\lbrack \omega ,f\theta ]_{\pi _{E}}=f[\omega ,\theta ]_{\pi _{E}}+%
\widetilde{\rho }_{E}(\omega )(f)\theta ,
\end{equation*}%
and it results that $(E^{\ast },[\cdot ,\cdot ]_{\pi _{E}},\widetilde{\rho }%
_{E})$ is a Lie algebroid, \cite{Kos}.

Moreover, we can define the Lichnerowicz type differential $d_{\pi
_{E}}:\Omega ^{p}(E)\rightarrow \Omega ^{p+1}(E)$ by 
\begin{eqnarray*}
d_{\pi _{E}}\omega (s_{1},\ldots ,s_{p+1}) &=&\sum_{i=1}^{p+1}(-1)^{i+1}%
\widetilde{\rho }_{E}(s_{i})\left( \omega (s_{1},\ldots ,\widehat{s_{i}}%
,\ldots ,s_{p+1})\right) \\
&&+\sum_{1\leq i<j\leq p+1}(-1)^{i+j}\omega \left( \lbrack s_{i},s_{j}]_{\pi
_{E}},s_{1},\ldots ,\widehat{s_{i}},\ldots ,\widehat{s_{j}},\ldots
,s_{p+1}\right) .
\end{eqnarray*}%
We have $d_{\pi _{E}}\circ d_{\pi _{E}}=0$ and, hence we get the cohomology
of Lie algebroid $(E^{\ast },[\cdot ,\cdot ]_{\pi _{E}},\widetilde{\rho }%
_{E})$ which generalize the Lichnerowicz-Poisson cohomology for Poisson
manifolds \cite{Li1, Va1}.

\section{Poisson structures on almost complex Lie algebroids}

\setcounter{equation}{0} In the first subsection of this section we briefly
present some elementary notions about almost complex Lie algebroids over
smooth manifolds. In the second subsection we introduce the notion of almost
complex Poisson structure on almost complex Lie algebroids and we describe
some examples of such structures. Also, the complex Lichnerowicz-Poisson
cohomology of almost complex Poisson Lie algebroids is introduced. In the
last subsection we consider the morphisms between almost complex Lie
algebroids preserving the almost complex Poisson structures and induce
almost complex Poisson structures on Lie subalgebroids. Also, we
characterize the morphisms of almost complex Poisson Lie algebroids in terms
of its Graph which is a $J_{E}$--coisotropic Lie subalgebroid of the
associated almost complex Poisson Lie algebroid given by direct product
structure.

\subsection{Almost complex Lie algebroids}

In \cite{B-R} is given the definition of an almost complex Lie algebroid
over an almost complex manifold, which generalizes the definition of an
almost complex Poisson manifold given in \cite{C-F-I-U}, as follows.

Consider a $2n$--dimensional almost complex manifold $M$, with an almost
complex structure $J_{M}:\Gamma (TM)\rightarrow \Gamma (TM)$ and $(E,\rho
_{E},[\cdot ,\cdot ]_{E})$ be a Lie algebroid over $M$ with $\mathrm{rank}%
\,E=2m$. {\ According to \cite{B-R}, an \textit{almost complex structure $%
J_{E}$} on $(E,\rho _{E},[\cdot ,\cdot ]_{E})$ is an endomorphism $%
J_{E}:\Gamma (E)\rightarrow \Gamma (E)$ such that $J_{E}^{2}=-\mathrm{%
id_{\Gamma (E)}}$ and $J_{M}\circ \rho _{E}=\rho _{E}\circ J_{E}$. However,
for our purpose we do not need the relation $J_{M}\circ \rho _{E}=\rho
_{E}\circ J_{E}$ and so we will consider the almost complex Lie algebroids
over smooth manifolds $M$. }

\begin{definition}
{\ A real Lie algebroid $(E,\rho _{E},[\cdot ,\cdot ]_{E})$ over a smooth
manifolds $M$ endowed with an almost complex endomorphism (i.e. an
endomorphism $J_{E}$ of }${E}${\ such that $J_{E}^{2}=-\mathrm{id_{\Gamma
(E)}}$) will be called an \textit{almost complex Lie algebroid}. }
\end{definition}

The standard examples of almost complex Lie algebroids are given by the
tangent bundle of an almost complex manifold $(M,J)$, the cotangent bundle
of an almost complex Poisson manifold $(M,J,\pi ^{2,0})$, the tangent bundle
of a complex foliation, the prolongation of a Lie algebroid over its vector
bundle projection, see \cite{I-Po}.

{By complexification of} the real vector bundle $E$ we obtain the complex
vector bundle $E_{\mathbb{C}}:=E\otimes_{\mathbb{R}}\mathbb{C}\rightarrow M$
and by extending the anchor map and the Lie bracket $\mathbb{C}$--linearly,
we obtain a complex Lie algebroid $(E_{\mathbb{C}},[\cdot,\cdot]_E,\rho_E)$
with the anchor map $\rho_E:\Gamma(E_{\mathbb{C}})\rightarrow\Gamma(TM_{%
\mathbb{C}})$, that is, a homomorphism of the complexified of corresponding
Lie algebras, and $[s_1,fs_2]_E=f[s_1,s_2]_E+\rho_E(s_1)(f)s_2$, for every $%
s_1,s_2\in\Gamma(E_{\mathbb{C}})$ and $f\in C^\infty(M)_{\mathbb{C}}=
C^{\infty}(M)\otimes_{\mathbb{R}}\mathbb{C}$. Also, extending $\mathbb{C}$%
--linearly the almost complex structure $J_E$, we obtain the almost complex
structure $J_E$ on $E_{\mathbb{C}}$.

As usual, we have a splitting 
\begin{equation*}
E_{\mathbb{C}}=E^{1,0}\oplus E^{0,1}
\end{equation*}
according to the eigenvalues $\pm i$ of $J_E$ on $E_{\mathbb{C}}$. We also
have 
\begin{equation}  \label{II1}
\Gamma(E^{1,0})=\{s-iJ_Es\,|\,s\in\Gamma(E)\}\,,\,\Gamma(E^{0,1})=\{s+iJ_Es%
\,|\,s\in\Gamma(E)\}.
\end{equation}
Similarly, we have the splitting 
\begin{equation*}
E^*_{\mathbb{C}}:=E^*\otimes_{\mathbb{R}}\mathbb{C}=(E^{1,0})^*\oplus
(E^{0,1})^*
\end{equation*}
according to the eigenvalues $\pm i$ of $J^*_E$ on $E^*_{\mathbb{C}}$, where 
$J_E^*$ is the natural almost complex structure induced on $E^*$. We also
have 
\begin{equation}  \label{II2}
\Gamma((E^{1,0})^*)=\{\omega-iJ^*_E\omega\,|\,\omega\in\Gamma(E^*)\}\,,\,%
\Gamma((E^{0,1})^*)=\{\omega+iJ^*_E\omega\,|\,\omega\in\Gamma(E^*)\}.
\end{equation}
We set 
\begin{equation*}
\bigwedge^{p,q}(E)=\bigwedge^p(E^{1,0})^*\wedge\bigwedge^q(E^{0,1})^* \,\,%
\mathrm{and}\,\,\Omega^{p,q}(E)=\Gamma\left(\bigwedge^{p,q}(E)\right).
\end{equation*}
Then, the differential $d_E$ of the complex $\Omega^{\bullet}(E)=%
\bigoplus_{p,q}\Omega^{p,q}(E)$ splits into the sum 
\begin{equation*}
d_E=\partial^\prime_E+\partial_E+\overline{\partial}_E+\partial^{\prime%
\prime}_E,
\end{equation*}
where 
\begin{equation*}
\partial^\prime_E:\Omega^{p,q}(E)\rightarrow\Omega^{p+2,q-1}(E)\,,\,%
\partial_E:\Omega^{p,q}(E)\rightarrow\Omega^{p+1,q}(E),
\end{equation*}
\begin{equation*}
\overline{\partial}_E:\Omega^{p,q}(E)\rightarrow\Omega^{p,q+1}(E)\,,\,%
\partial^{\prime\prime}_E:\Omega^{p,q}(E)\rightarrow\Omega^{p-1,q+2}(E).
\end{equation*}

For an almost complex Lie algebroid $(E,\rho_E,[\cdot,\cdot]_E, J_E)$ we can
consider the Nijenhuis tensor of $J_E$ defined by 
\begin{equation}  \label{2.1}
N_{J_E}(s_1,s_2)=[J_Es_1,J_Es_2]_E-J_E[s_1,J_Es_2]_E-J_E[J_Es_1,s_2]_E-[s_1,s_2]_E\,,\,\forall\,s_1,s_2\in\Gamma(E).
\end{equation}

\begin{definition}
The almost complex structure $J_E$ of almost complex Lie algebroid $%
(E,\rho_E,[\cdot,\cdot]_E, J_E)$ is called \textit{integrable} if $N_{J_E}=0$%
.
\end{definition}

Also, using a standard procedure from almost complex geometry, \cite{G-O,Ya}%
, we have proved the following Newlander-Nirenberg type theorem:

\begin{theorem}
\label{t1} \cite{I-Po} For an almost complex Lie algebroid $%
(E,\rho_E,[\cdot,\cdot]_E, J_E)$ over a smooth manifold $M$ the following
assertions are equivalent:

\begin{enumerate}
\item[(i)] If $s_1,s_2\in\Gamma(E^{1,0})$ then $[s_1,s_2]_E\in%
\Gamma(E^{1,0}) $;

\item[(ii))] If $s_1,s_2\in\Gamma(E^{0,1})$ then $[s_1,s_2]_E\in%
\Gamma(E^{0,1})$;

\item[(iii)] $d_E\Omega^{1,0}(E)\subset \Omega^{2,0}(E)+\Omega^{1,1}(E)$ and 
$d_E\Omega^{0,1}(E)\subset \Omega^{1,1}(E)\oplus\Omega^{0,2}(E)$;

\item[(iv)] $d_E\Omega^{p,q}(E)\subset \Omega^{p+1,q}(E)+\Omega^{p,q+1}(E)$;

\item[(v)] the real Nijenhuis $N_{J_E}$ from \eqref{2.1} vanish, namely $J_E$
is integrable.
\end{enumerate}
\end{theorem}

The above Newlander-Nirenberg type theorem says that for any integrable
almost complex structure $J_E$ on an almost complex Lie algebroid $%
(E,\rho_E,[\cdot,\cdot]_E)$ we have the usual decomposition 
\begin{equation*}
d_E=\partial_E+\overline{\partial}_E.
\end{equation*}
From $d_E^2=d_E\circ d_E=0$ we obtain the following identities: 
\begin{equation}  \label{II5}
\partial_E^2=\overline{\partial}_E^2=\partial_E\overline{\partial}_E+%
\overline{\partial}_E\partial_E=0.
\end{equation}
Hence, in this case we have a Dolbeault type Lie algebroid cohomology as the
cohomology of the complex $(\Omega^{p,\bullet}(E),\overline{\partial}_E)$.

We notice that in more situations in this paper we will consider the case
when $J_E$ is integrable.

Let us consider now $(E,\rho_E,[\cdot,\cdot]_E, J_E)$ and $%
(E^\prime,\rho_{E^\prime},[\cdot,\cdot]_{E^\prime},J_{E^\prime})$ be two
almost complex Lie algebroids over a smooth manifold $M$.

A $M$--morphism $\varphi$ of almost complex Lie algebroids is naturally
extended by $\mathbb{C}$--linearity to $\varphi:(E_{\mathbb{C}%
},\rho_E,J_E)\rightarrow(E^\prime_{\mathbb{C}},\rho_{E^\prime},J_{E^\prime})$
over $M$ and it is called \textit{almost complex} if 
\begin{equation}  \label{II6}
\varphi\circ J_E=J_{E^\prime}\circ\varphi.
\end{equation}

\begin{proposition}
\cite{I-Po} If $\varphi:(E,\rho_E,J_E)\rightarrow(E^\prime,\rho_{E^%
\prime},J_{E^\prime})$ is a $M$--morphism of almost complex Lie algebroids
over $M$, then the following assertions are equivalent:

\begin{enumerate}
\item[(i)] If $s_1\in\Gamma(E^{1,0})$ then $\varphi(s_1)\in\Gamma(E^{\prime
1,0})$;

\item[(ii)] If $s_1\in\Gamma(E^{0,1})$ then $\varphi(s_1)\in\Gamma(E^{\prime
0,1})$;

\item[(iii)] If $\omega^{\prime}\in\Omega^{p,q}(E^{^{\prime }})$ then $%
\varphi^*\omega^{\prime}\in\Omega^{p,q}(E)$, where 
\begin{equation*}
(\varphi^*\omega^\prime)(s_1,\ldots,s_p,t_1,\ldots,t_q)=\omega^\prime(%
\varphi(s_1),\ldots,\varphi(s_p),\varphi(t_1),\ldots,\varphi(t_q))
\end{equation*}
for any $s_1,\ldots,s_p\in\Gamma(E^{1,0})$ and $t_1,\ldots,t_q\in%
\Gamma(E^{0,1})$.

\item[(iv)] The morphism $\varphi$ is almost complex.
\end{enumerate}
\end{proposition}

\subsection{Almost complex Poisson Lie algebroids}

Let $(E,\rho_E,[\cdot,\cdot]_E, J_E)$ be an almost complex Lie algebroid
over a smooth manifold $M$. Then the complexification $\mathcal{V}^\bullet_{%
\mathbb{C}}(E)=\mathcal{V}^\bullet(E)\otimes_{\mathbb{R}}\mathbb{C}$ of the
Grassmann algebra of multisections of $E$, admits the decomposition $%
\mathcal{V}_{\mathbb{C}}^{p+q}(E)=\bigoplus_{p,q}\mathcal{V}^{p,q}(E)$,
where we have put $\mathcal{V}^{p,q}(E)=\Gamma\left(\bigwedge^pE^{1,0}\wedge%
\bigwedge^qE^{0,1}\right)$.

As in the usual case of almost complex Poisson manifolds, \cite{C-F-I-U},
with respect to {the graduation above,} we have the following decomposition
of the Schouten-Nijenhuis bracket on $E$: 
\begin{equation}
\lbrack S,T]_{E}\in \mathcal{V}^{p+r-2,q+s+1}(E)\oplus \mathcal{V}%
^{p+r-1,q+s}(E)\oplus \mathcal{V}^{p+r,q+s-1}(E)\oplus \mathcal{V}%
^{p+r+1,q+s-2}(E),  \label{III1}
\end{equation}%
for every $S\in \mathcal{V}^{p,q}(E)$ and $T\in \mathcal{V}^{r,s}(E)$. In
particular, the bracket of two sections of type $(1,0)$ may not be of type $%
(1,0)$.

If the almost complex structure $J_E$ is integrable then the
Schouten-Nijenhuis bracket on $E$ has the property 
\begin{equation}  \label{III2}
[S,T]_{E}\in\mathcal{V}^{p+r-1,q+s}(E)\oplus\mathcal{V}^{p+r,q+s-1}(E)
\end{equation}
for every $S\in\mathcal{V}^{p,q}(E)$ and $T\in\mathcal{V}^{r,s}(E)$. In
particular, the bracket of two sections of type $(1,0)$ is again of type $%
(1,0)$.

\begin{definition}
Let $(E,\rho_E,[\cdot,\cdot]_E, J_E)$ be an almost complex Lie algebroid.
Then a $(2,0)$--section $\pi^{2,0}_E\in\mathcal{V}^{2,0}(E)$ is called an 
\textit{almost complex Poisson structure} if 
\begin{equation}  \label{III3}
\left[\pi^{2,0}_E,\pi^{2,0}_E\right]_E=0\,\,\mathrm{and}\,\,\left[%
\pi^{2,0}_E,\overline{\pi^{2,0}_E}\right]_E=0.
\end{equation}
\end{definition}

Associated to an almost complex Poisson structure on the almost complex Lie
algebroid $(E,\rho _{E},[\cdot ,\cdot ]_{E},J_{E})$ there is a bracket on
the algebra of complex functions $C^{\infty }(M)_{\mathbb{C}}=C^{\infty
}(M)\otimes _{\mathbb{R}}\mathbb{C}$, 
\begin{equation*}
\{\cdot ,\cdot \}:C^{\infty }(M)_{\mathbb{C}}\times C^{\infty }(M)_{\mathbb{C%
}}\rightarrow C^{\infty }(M)_{\mathbb{C}}
\end{equation*}%
given by $\{f,g\}=\imath _{\pi _{E}^{2,0}}(d_{E}f\wedge d_{E}g)$; and it
satisfies the {skew-symmetry condition}, the Leibniz rule and the Jacobi
identity (which in fact is equivalent to the first relation from \eqref{III3}%
).

Also, from \eqref{III1}, we easily obtain

\begin{proposition}
If $\pi^{2,0}_E$ defines an almost complex Poisson structure on an almost
complex Lie algebroid $(E,\rho_E,[\cdot,\cdot]_E, J_E)$, then $%
\pi_E=\pi^{2,0}_E+\overline{\pi^{2,0}_E}$ and $\pi^{\prime}_E=i\left(%
\pi^{2,0}_E-\overline{\pi^{2,0}_E}\right)$ are real Poisson structures on $E$%
; the converse also holds.
\end{proposition}

\begin{remark}
The converse in the above proposition is not true if only one of the tensors 
$\pi_E$ or $\pi^{\prime}_E$ is real Poisson, unless $J_E$ is integrable.
\end{remark}

\begin{example}
\label{e2.1} ($\partial _{E}$--holomorphic symplectic structures). According
to \cite{I-M-D-M-P} a symplectic structure on a Lie algebroid $(E,\rho
_{E},[\cdot ,\cdot ]_{E})$ is defined by a non-degenerated $d_{E}$--closed $%
2 $--form $\omega \in \Omega ^{2}(E)$. For instance it can be induced by a K%
\"{a}hlerian structure $g$ on an almost complex Lie algebroid $(E,\rho
_{E},[\cdot ,\cdot ]_{E},J_{E})$, see \cite{I-Po}. As similar to \cite%
{Pa1,Pa2}, when $J_{E}$ is integrable, we can consider the notion of a $%
\partial _{E}$--symplectic structure on an almost complex Lie algebroid $%
(E,\rho _{E},[\cdot ,\cdot ]_{E},J_{E})$ as a non-degenerated $\partial _{E}$%
--closed $(2,0)$--form $\omega ^{2,0}\in \Omega ^{2,0}(E)$. As it happens in
the case of real symplectic structures on Lie algebroids, there exists the
isomorphism $\mu :\Gamma (E^{1,0})\rightarrow \Omega ^{1,0}(E)$, defined by $%
\mu (s)=\imath _{s}\omega ^{2,0}$. Extending $\mu $ to a mapping on the
associated Grassmann algebras, we consider the $(2,0)$-section $\pi
_{E}^{2,0}=-\mu ^{-1}(\omega ^{2,0})\in \mathcal{V}^{2,0}(E)$. Equivalently, 
$\pi _{E}^{2,0}$ is defined by $\pi _{E}^{2,0}(\partial _{E}f,\partial
_{E}g)=-\omega ^{2,0}(s_{f},s_{g})$, where $s_{f}$ is the Hamiltonian
section of $f\in C^{\infty }(M)_{\mathbb{C}}$ defined as the unique section
of type $(1,0)$ on $E$ such that $\imath _{s_{f}}\omega ^{2,0}=\partial
_{E}f $. In order to prove that $\pi _{E}^{2,0}$ defines an almost complex
Poisson structure on the almost complex Lie algebroid $(E,\rho _{E},[\cdot
,\cdot ]_{E},J_{E})$, we have 
\begin{eqnarray*}
\imath _{\lbrack s_{f},s_{g}]_{E}}\omega ^{2,0} &=&\mathcal{L}_{s_{f}}\imath
_{s_{g}}\omega ^{2,0}-\imath _{s_{g}}\mathcal{L}_{s_{f}}\omega ^{2,0} \\
&=&\imath _{s_{f}}d_{E}(\partial _{E}g)+d_{E}\imath _{s_{f}}\imath
_{s_{g}}\omega ^{2,0}-\imath _{s_{g}}d_{E}(\partial _{E}f),
\end{eqnarray*}%
and comparing the {bigrade} of both terms of above equality, we have

\begin{enumerate}
\item[(i)] $[s_f,s_g]_E=s_{\{f,g\}}$;

\item[(ii)] $\imath_{s_f}\overline{\partial}_E\partial_Eg-\imath_{s_g}%
\overline{\partial}_E\partial_Ef-\overline{\partial}_E\{f,g\}=0$.
\end{enumerate}

Now, as a consequence of (i) and that {\ $\omega ^{2,0}$ is $\partial _{E}$%
--closed }, we obtain the Jacobi identity, hence $\left[ \pi _{E}^{2,0},\pi
_{E}^{2,0}\right] _{E}=0$. If in addition, we require that the base manifold 
$M$ to be complex and $\omega ^{2,0}$ to be a holomorphic $2$--form on $E$
(that is a $(2,0)$--form on $E$ with coefficients holomorphic functions on $%
M $), then $\pi _{E}^{2,0}$ is a holomorphic $2$--section (or equivalently $%
\overline{\pi _{E}^{2,0}}$ is anti-holomorphic $2 $--section) and then, the
condition $\left[ \pi _{E}^{2,0},\overline{\pi _{E}^{2,0}}\right] _{E}=0$ is
trivially satisfied, and so $\pi _{E}^{2,0}$ defines an almost complex
Poisson structure on $E$.
\end{example}

\begin{remark}
If we consider in above example the particular case of the almost complex
Lie algebroid $\left(T\mathcal{F},i_{\mathcal{F}},[\cdot,\cdot]_{\mathcal{F}%
},J_{\mathcal{F}}\right)$ associated to a complex foliation $\mathcal{F}$,
see \cite{I-Po}, then similarly to Proposition 6.1 from \cite{C-F-I-U}, we
obtain an almost complex Poisson structure $\pi_{\mathcal{F}}^{2,0}$ on $T%
\mathcal{F}$ from a foliated complex symplectic $2$--form $\omega_{\mathcal{F%
}}$, that is, $\omega_{\mathcal{F}}$ is a closed foliated $\mathcal{F}$%
--holomorphic $(2,0)$--form.
\end{remark}

\begin{example}
(Complete lift to the prolongation of a Lie algebroid). For a Lie algebroid $%
(E,\rho_E,[\cdot,\cdot]_E)$ with $\mathrm{rank}\,E=m$ we can consider the
prolongation of $E$ over its vector bundle projection, see {\ \cite{H-M,
Ma1, Po}}, which is a vector bundle $p_L:\mathcal{L}^p(E)\rightarrow E$ of $%
\mathrm{rank}\,\mathcal{L}^p(E)=2m$ which has a Lie algebroid structure over 
$E$. More exactly, $\mathcal{L}^p(E)$ is the subset of $E\times TE$ defined
by $\mathcal{L}^p(E)=\{(u,z)\,|\,\rho_E(u)=p_*(z)\}$, where $%
p_*:TE\rightarrow TM $ is the canonical projection. The projection on the
second factor $\rho_{\mathcal{L}^p(E)}:\mathcal{L}^p(E)\rightarrow TE$,
given by $\rho_{\mathcal{L}^p(E)}(u,z)=z$ will be the anchor of the
prolongation Lie algebroid $\left(\mathcal{L}^p(E),\rho_{\mathcal{L}%
^p(E)},[\cdot,\cdot]_{\mathcal{L}^p(E)}\right)$ over $E$. According to {\ 
\cite{Ma1, Po}}, we can consider the \textit{vertical lift} $s^v$ and the 
\textit{complete lift} $s^c$ of a section $s\in\Gamma(E)$ as sections of $%
\mathcal{L}^p(E)$ as follows. The local basis of $\Gamma(\mathcal{L}^p(E))$
is given by $\left\{\mathcal{X}_a(u)=\left(e_a(p(u)),\rho_a^{i}\frac{\partial%
}{\partial x^{i}}|_u\right),\mathcal{V}_a=\left(0,\frac{\partial}{\partial
y^{a}}\right)\right\}$, where $\left\{\frac{\partial}{\partial x^{i}},\frac{%
\partial}{\partial y^{a}}\right\}$, $i=1,\ldots,n=\dim M$, $a=1\ldots,m=%
\mathrm{rank}\,E$, is the local basis on $TE$. Then, the vertical and
complete lifts, respectively, of a section $s=s^{a}e_a\in\Gamma(E)$ are
given by 
\begin{equation*}
s^v=s^{a}\mathcal{V}_a\,,\,s^c=s^{a}\mathcal{X}_a+\left(%
\rho_E(e_c)(s^{a})-C^{a}_{bc}s^b\right)y^c\mathcal{V}_a.
\end{equation*}
In particular, $e_a^v=\mathcal{V}_a$ and $e_a^c=\mathcal{X}_a-C^b_{ac}y^c%
\mathcal{V}_b$.

Now, if $\mathrm{rank}\,E=2m$ and $J_E$ is an almost complex structure on
the Lie algebroid $(E,\rho_E,[\cdot,\cdot]_E)$, then $J_E^c$ is an almost
complex structure on $\left(\mathcal{L}^p(E),\rho_{\mathcal{L}%
^p(E)},[\cdot,\cdot]_{\mathcal{L}^p(E)}\right)$ (because one of the
properties of the complete lift is: for $p(T)$ a polynomial, then $%
p(T^c)=p(T)^c$). In fact, if $J_E$ is expressed locally by $%
J_E=J_b^{a}e_a\otimes e^b$, then $J_E^c$ is locally given by 
\begin{equation*}
J_E^c=\left(\rho_E(e_c)(J^{a}_b)-C^{a}_{db}J^{d}_c\right)y^c\mathcal{V}%
_a\otimes\mathcal{X}^{b}+J^{a}_b\mathcal{X}_a\otimes\mathcal{X}^{b}+J^{a}_b%
\mathcal{V}_a\otimes\mathcal{V}^b,
\end{equation*}
where $\{\mathcal{X}^{a},\mathcal{V}^b\}$ is the dual basis of $\{\mathcal{X}%
_a,\mathcal{V}_b\}$.

Moreover, $J_{E}^{c}s^{v}=(J_{E}s)^{v}$ and $J_{E}^{c}s^{c}=(J_{E}s)^{c}$,
so we have that if $s\in \Gamma (E^{1,0})$ (or $s\in \Gamma (E^{0,1})$),
then $s^{v},s^{c}\in \Gamma (\mathcal{L}^{p}(E)^{1,0})$ (or $s^{v},s^{c}\in
\Gamma (\mathcal{L}^{p}(E)^{0,1})$). Taking into account now that 
\begin{equation}
(S\wedge T)^{v}=S^{v}\wedge T^{v}\,\,\mathrm{and}\,\,(S\wedge
T)^{c}=S^{c}\wedge T^{v}+S^{v}\wedge T^{c},  \label{III4}
\end{equation}%
for $S,T\in \mathcal{V}^{\bullet }(E)$, it follows that the {bigraduation
with} respect to the complex structures $J_{E}$ and $J_{E}^{c}$ is preserved
when considering the vertical and complete lifts of multisections, that is,
if $S\in \mathcal{V}^{p,q}(E)$, then $S^{v},S^{c}\in \mathcal{V}^{p,q}(%
\mathcal{L}^{p}(E))$. Furthermore, since $N_{J_{E}^{c}}=\left(
N_{J_{E}}\right) ^{c}$, then $J_{E}^{c}$ is integrable if and only if $J_{E}$
is integrable.

Now, from \eqref{III4} and 
\begin{equation*}
[s_1^v,s_2^v]_{\mathcal{L}^p(E)}=0\,,\,[s_1^v,s_2^c]_{\mathcal{L}%
^p(E)}=[s_1,s_2]^v_{E}\,,\,[s_1^c,s_2^c]_{\mathcal{L}^p(E)}=[s_1,s_2]^c_{E},
\end{equation*}
see \cite{Ma}, it follows that for the Schouten-Nijenhuis bracket we have 
\begin{equation*}
[S,T]_E^c=[S^c,T^c]_{\mathcal{L}^p(E)},
\end{equation*}
for $S,T\in\mathcal{V}^\bullet(E)$. Hence, if $\left(J_E,\pi^{2,0}_E\right)$
is an almost complex Poisson structure on $E$, then $\left(J_E^c,(%
\pi^{2,0}_E)^c\right)$ is an almost complex Poisson structure on $\mathcal{L}%
^p(E)$.
\end{example}

\begin{example}
Let us consider again the prolongation Lie algebroid $\left(\mathcal{L}%
^p(E),\rho_{\mathcal{L}^p(E)},[\cdot,\cdot]_{\mathcal{L}^p(E)}\right)$ over $%
E$ from above example, and let $\nabla$ a linear connection on the Lie
algebroid $E$ (in particular a Riemannian Lie algebroid $(E,g)$ and the
Levi-Civita connection). Then, the connection $\nabla$ leads to a natural
decomposition of $\mathcal{L}^p(E)$ into vertical and horizontal subbundles 
\begin{equation*}
\mathcal{L}^p(E)=H\mathcal{L}^p(E)\oplus V\mathcal{L}^p(E),
\end{equation*}
where $V\mathcal{L}^p(E)=\mathrm{span}\,\{\mathcal{V}_a\}$ and $H\mathcal{L}%
^p(E)=\mathrm{span}\,\{\mathcal{H}_a=\mathcal{X}_a-\Gamma^b_{ac}y^c\mathcal{V%
}_b\}$, where $\Gamma^b_{ac}(x)$ are the local coefficients of the linear
connection $\nabla$. We notice that, the above decomposition can be obtained
also by a nonlinear (Ehresmann) connection.

The horizontal lift $s^h$ of a section $s=s^{a}e_a\in\Gamma(E)$ to $\mathcal{%
L}^p(E)$ is locally given by 
\begin{equation*}
s^h=s^{a}\mathcal{H}_a=s^{a}(\mathcal{X}_a-\Gamma^b_{ac}y^c\mathcal{V}_b).
\end{equation*}
Then, there exists an almost complex structure $J^1_{\mathcal{L}^p(E)}$ on $%
\mathcal{L}^p(E)$, induced by the connection $\nabla$, given by 
\begin{equation*}
J^1_{\mathcal{L}^p(E)}(s^h)=s^v\,,\,J^1_{\mathcal{L}^p(E)}(s^v)=-s^h.
\end{equation*}
A simple example of almost complex Poisson structure on $\mathcal{L}^p(E)$
with respect to the almost complex structure $J^1_{\mathcal{L}^p(E)}$ can be
constructed as follows: Let $s_1,s_2\in\Gamma(E)$, then $s_1^h-is_1^v$ and $%
s_2^h-is_2^v$ are complex sections of type $(1,0)$ on $\mathcal{L}^p(E)$,
and we consider 
\begin{eqnarray*}
\pi^{2,0}_{\mathcal{L}^p(E)}&=&(s_1^h-is_1^v)\wedge(s_2^h-is_2^v) \\
&=&s_1^h\wedge s_2^h-i(s_1^h\wedge s_2^v+s_1^v\wedge s_2^h)-s_1^v\wedge s_2^v
\\
&=&s_1^h\wedge s_2^h-i(s_1\wedge s_2)^h-(s_1\wedge s_2)^v.
\end{eqnarray*}
Taking into account that 
\begin{equation*}
[s_1^h,s_2^h]_{\mathcal{L}^p(E)}=[s_1,s_2]_E^h+(R(s_1,s_2)u)^v\,,%
\,[s_1^h,s_2^v]_{\mathcal{L}^p(E)}=\left(\nabla_{s_1}s_2\right)^v\,,%
\,[s_1^v,s_2^v]_{\mathcal{L}^p(E)}=0,
\end{equation*}
it follows that $\pi^{2,0}_{\mathcal{L}^p(E)}$ is an almost complex Poisson
structure on $\mathcal{L}^p(E)$ if $[s_1,s_2]_E=0$, $\nabla_{s_1}s_2=0$, $%
R(s_1,s_2)=0$ and $T(s_1,s_2)=0$ (in particular, if $\nabla$ is the
Levi-Civita connection of a Riemannian metric on $E$, then $T=0$ always).
Here, the curvature $R$ satisfies $R=-N_h$, where $N_h$ is the Nijenhuis
tensor of the horizontal projection.

If, moreover, $\mathrm{rank}\,E=2m$ and $J_{E}$ is an almost complex
structure on $E$, then we can consider its horizontal lift $J_{E}^{h}$ on $%
\mathcal{L}^{p}(E)$, which is again an almost complex structure, such that $%
J_{E}^{h}(s^{v})=(J_{E}s)^{v}$ and $J_{E}^{h}(s^{h})=(J_{E}s)^{h}$. It is
easy to see that the bigraduations with respect to $J_{E}$ and $J_{E}^{h}$
is preserved by vertical and horizontal lifts and hence, the determination
of almost complex Poisson structures on $\mathcal{L}^{p}(E)$ with respect to
the almost complex structure $J_{E}^{h}$ can be of some {interest}.
\end{example}

\begin{example}
\label{e3.4} (Direct product structure). The direct product of two Lie
algebroids $\left( E_{1},\rho _{E_{1}},[\cdot ,\cdot ]_{E_{1}}\right) $ over 
$M_{1}$ and $\left( E_{2},\rho _{E_{2}},[\cdot ,\cdot ]_{E_{2}}\right) $
over $M_{2}$ is defined in {\cite[pg. 155]{Mack2}}, as a Lie algebroid
structure $E_{1}\times E_{2}\rightarrow M_{1}\times M_{2}$. Let us briefly
recall this construction. The general sections of $E_{1}\times E_{2}$ are of
the form $s=\sum (f_{i}\otimes s_{i}^{1})\oplus \sum (g_{j}\otimes
s_{j}^{2}) $, where $f_{i},g_{j}\in C^{\infty }(M_{1}\times M_{2})$, $%
s_{i}^{1}\in \Gamma (E_{1})$, $s_{j}^{2}\in \Gamma (E_{2})$, and the anchor
map is defined by 
\begin{equation*}
\rho _{E}\left( \sum (f_{i}\otimes s_{i}^{1})\oplus \sum (g_{j}\otimes
s_{j}^{2})\right) =\sum (f_{i}\otimes \rho _{E_{1}}(s_{i}^{1}))\oplus \sum
(g_{j}\otimes \rho _{E_{2}}(s_{j}^{2})).
\end{equation*}%
Imposing the conditions 
\begin{equation*}
\lbrack 1\otimes s^{1},1\otimes t^{1}]_{E}=1\otimes \lbrack
s^{1},t^{1}]_{E_{1}}\,,\,[1\otimes s^{1},1\otimes t^{2}]_{E}=0,
\end{equation*}%
\begin{equation*}
\lbrack 1\otimes s^{2},1\otimes t^{2}]_{E}=1\otimes \lbrack
s^{2},t^{2}]_{E_{2}}\,,\,[1\otimes s^{2},1\otimes t^{1}]_{E}=0,
\end{equation*}%
for every $s^{1},t^{1}\in \Gamma (E_{1})$ and $s^{2},t^{2}\in \Gamma (E_{2})$%
, it follows that for $s=\sum (f_{i}\otimes s_{i}^{1})\oplus \sum
(g_{j}\otimes s_{j}^{2})$ and $s^{\prime }=\sum (f_{k}^{\prime }\otimes
s_{k}^{\prime 1})\oplus \sum (g_{l}^{\prime }\otimes s_{l}^{\prime 2})$, we
have, using Leibniz condition, the following expression for bracket on $%
E=E_{1}\times E_{2}$: 
\begin{eqnarray*}
\lbrack s,s^{\prime }]_{E} &=&\left( \sum f_{i}f_{k}^{\prime }\otimes
\lbrack s_{i}^{1},s_{k}^{\prime 1}]_{E_{1}}+\sum f_{i}\rho
_{E_{1}}(s_{i}^{1})(f_{k}^{\prime })\otimes s_{k}^{\prime 1}-\sum
f_{k}^{\prime }\rho _{E_{1}}(s_{k}^{\prime 1})(f_{i})\otimes s_{i}^{1}\right)
\\
&&\oplus \left( \sum g_{j}g_{l}^{\prime }\otimes \lbrack
s_{j}^{2},s_{l}^{\prime 2}]_{E_{2}}+\sum g_{j}\rho
_{E_{2}}(s_{j}^{2})(g_{l}^{\prime })\otimes s_{l}^{\prime 2}-\sum
g_{l}^{\prime }\rho _{E_{2}}(s_{l}^{\prime 2})(g_{j})\otimes
s_{j}^{2}\right) .
\end{eqnarray*}%
Now, we consider $E_{1}$ and $E_{2}$ endowed with almost complex Poisson
structures $\left( J_{E_{1}},\pi _{E_{1}}^{2,0}\right) $ and $\left(
J_{E_{2}},\pi _{E_{2}}^{2,0}\right) $, respectively. We define the almost
complex structure $J_{E}$ on $E=E_{1}\times E_{2}$ by $J_{E}(s)=\sum
(f_{i}\otimes J_{E_{1}}(s_{i}^{1}))\oplus \sum (g_{j}\otimes
J_{E_{2}}(s_{j}^{2}))$ and the $2$--section $\pi _{E}^{2,0}=\pi
_{E_{1}}^{2,0}+\pi _{E_{2}}^{2,0}$. Since $[s^{1},s^{2}]_{E}=0$ for every $%
s^{1}\in \Gamma (E_{1})$, $s^{2}\in \Gamma (E_{2})$, it follows that $\pi
_{E}^{2,0}$ is an almost complex Poisson structure on the direct product $%
E_{1}\times E_{2}$.
\end{example}

Let us consider now $\pi^{2,0}_E$ be an almost complex Poisson structure on
the almost complex Lie algebroid $\left(E,\rho_{E},[\cdot,\cdot]_{E},
J_E\right)$. It is well known that, associated to the real Poisson structure 
$\pi_E=\pi^{2,0}_E+\overline{\pi^{2,0}_E}$, there is a morphism 
\begin{equation*}
\pi_E^\#:\Omega^1(E)\rightarrow\mathcal{V}^1(E),
\end{equation*}
given by $\pi^\#_E(\varphi)(\psi)=\pi_E(\varphi,\psi)$ for $%
\varphi,\psi\in\Omega^1(E)$. Then, we can consider the complexified spaces
and extend $\pi^\#_E$ to the complex Grassmann algebras as follows: let $%
I^*_E$ be the adjoint of $I_E=\pi^\#_E$ (it is easy to see that $I_E^*=-I_E$%
), then we have 
\begin{equation*}
I_E:\Omega_{\mathbb{C}}^p(E)\rightarrow\mathcal{V}_{\mathbb{C}%
}^p(E)\,,\,I_E(\varphi)(\varphi_1,\ldots,\varphi_p)=\varphi(I_E^*\varphi_1,%
\ldots,I_E^*\varphi_p),
\end{equation*}
for $\varphi\in\Omega^p_{\mathbb{C}}(E)$ and $\varphi_1,\ldots,\varphi_p\in%
\Omega^1_{\mathbb{C}}(E)$.

For the $(2,0)$--section $\pi _{E}^{2,0}$ it can be defined similarly the
morphisms 
\begin{equation*}
\left( \pi _{E}^{2,0}\right) ^{\#}:\Omega ^{1,0}(E)\rightarrow \mathcal{V}%
^{1,0}(E)\,,\,\left( \overline{\pi _{E}^{2,0}}\right) ^{\#}:\Omega
^{0,1}(E)\rightarrow \mathcal{V}^{0,1}(E).
\end{equation*}%
Since $\pi _{E}^{\#}|_{\Omega ^{1,0}(E)}=\left( \pi _{E}^{2,0}\right) ^{\#}$%
, $\pi _{E}^{\#}|_{\Omega ^{0,1}(E)}=\left( \overline{\pi _{E}^{2,0}}\right)
^{\#}$ and $I_{E}$ is a morphism of Grassmann algebras, then $I_{E}$
preserves the bigraduation with respect to $J_{E}$.

Now, let $\pi^{2,0}_E$ be an almost complex Poisson structure on the almost
complex Lie algebroid $\left(E,\rho_{E},[\cdot,\cdot]_{E}, J_E\right)$ such
that $J_E$ is integrable. We define the following operators

\begin{enumerate}
\item[(i)] $\sigma_E:\mathcal{V}_{\mathbb{C}}^p(E)\rightarrow\mathcal{V}_{%
\mathbb{C}}^{p+1}(E)$, by $\sigma_E(S)=-\left[S,\pi_E\right]_E$, for $S\in%
\mathcal{V}_{\mathbb{C}}^p(E)$;

\item[(ii)] $\sigma_E^1:\mathcal{V}^{p,q}(E)\rightarrow\mathcal{V}%
^{p+1,q}(E)\oplus\mathcal{V}^{p+2,q-1}(E)$, by $\sigma_E^1(S)=-\left[%
S,\pi^{2,0}_E\right]_E$, for $S\in\mathcal{V}^{p,q}(E)$;

\item[(ii)] $\sigma_E^2:\mathcal{V}^{p,q}(E)\rightarrow\mathcal{V}%
^{p-1,q+2}(E)\oplus\mathcal{V}^{p,q+1}(E)$, by $\sigma_E^2(S)=-\left[S,%
\overline{\pi^{2,0}_E}\right]_E$, for $S\in\mathcal{V}^{p,q}(E)$.
\end{enumerate}

Clearly, $\overline{\sigma_E^1(S)}=\sigma_E^2(\overline{S})$, for $S\in%
\mathcal{V}^{p,q}(E)$. Also, in the same way we can define $%
\sigma_E,\sigma_E^1$ and $\sigma_E^2$ in the case when $J_E$ is not
integrable, but the image space has more components.

If $J_E$ is integrable, as usual, we have $I_E(d_E\varphi)=-\sigma_E(I_E%
\varphi)$ and for $\varphi\in\Omega^{p,q}(E)$ it follows

\begin{enumerate}
\item[(i)] $I_E(d_E\varphi)=I_E(\partial_E\varphi)+I_E(\overline{\partial}%
_E\varphi)\in\mathcal{V}^{p+1,q}(E)\oplus\mathcal{V}^{p,q+1}(E)$;

\item[(ii)] $\sigma_E(I_E\varphi)=\sigma_E^1(I_E\varphi)+\sigma_E^2(I_E%
\varphi)\in\mathcal{V}^{p+1,q}(E)\oplus\mathcal{V}^{p+2,q-1}(E)\oplus%
\mathcal{V}^{p,q+1}(E)\oplus\mathcal{V}^{p-1,q+2}(E)$.
\end{enumerate}

\begin{corollary}
Let $\pi^{2,0}_E$ be an almost complex {Poisson} structure on $%
\left(E,\rho_{E},[\cdot,\cdot]_{E}, J_E\right)$. Suppose that $J_E$ is
integrable. Then 
\begin{equation*}
I_E(\partial_E\varphi)=-\sigma_E^1(I_E\varphi)\in\mathcal{V}%
^{p+1,q}(E)\,,\,I_E(\overline{\partial}_E\varphi)=-\sigma_E^2(I_E\varphi)\in%
\mathcal{V}^{p,q+1}(E),
\end{equation*}
for every $\varphi\in\Omega^{p,q}(E)$.
\end{corollary}

\begin{proposition}
\label{p3.2} A nondegenerate almost complex Poisson structure $\pi^{2,0}_E$
on an almost complex Lie algebroid $\left(E,\rho_{E},[\cdot,\cdot]_{E},
J_E\right)$ with $J_E$ integrable defines a $\partial_E$--symplectic
structure on $E$.
\end{proposition}

\proof If $\pi _{E}^{2,0}$ is {a} nondegenerate almost complex Poisson
structure, then $\pi _{E}=\pi _{E}^{2,0}+\overline{\pi _{E}^{2,0}}$ is also
nondegenerate and the morphism $I_{E}$ is an isomorphism. Now, we consider $%
\omega ^{2,0}=I_{E}^{-1}\left( \pi _{E}^{2,0}\right) $ and, then 
\begin{equation*}
I_{E}\left( \partial _{E}\omega ^{2,0}\right) =-\sigma _{E}^{1}\left(
I_{E}\omega ^{2,0}\right) =-\sigma _{E}^{1}\left( \pi _{E}^{2,0}\right) = 
\left[ \pi _{E}^{2,0},\pi _{E}^{2,0}\right] _{E}=0,
\end{equation*}%
and, since $I_{E}$ is an isomorphism, it follows that $\omega ^{2,0}$
defines a $\partial _{E}$--symplectic structure on $E$. \qed

For a complex smooth function $f\in C^{\infty }(M)_{\mathbb{C}}$, there is
an associated complex section of type $(1,0)$ of $E$ (called the \textit{%
Hamiltonian section} associated to $f$, see also Example \ref{e2.1}) defined
by 
\begin{equation*}
s_{f}=\imath _{\partial _{E}f}\pi _{E}^{2,0}=\pi _{E}^{2,0}(\partial _{E}f).
\end{equation*}%
From the Jacobi identity of the bracket $\{\cdot ,\cdot \}$ associated to $%
\pi _{E}^{2,0}$, we obtain $[s_{f},s_{g}]_{E}=s_{\{f,g\}}$ (see also Example %
\ref{e2.1}); hence, the space $\mathrm{Ham}\,_{\pi _{E}^{2,0}}$ of
Hamiltonian $(1,0)$--sections on $E$ is a subspace of $\mathcal{V}^{1,0}(E)$
with a Lie algebra structure respect to the Lie bracket of sections of $E$,
which in the case when $J_{E}$ is integrable is actually a Lie subalgebra of 
$\mathcal{V}^{1,0}(E)$. Moreover, we have

\begin{proposition}
Every Hamiltonian section of $\pi^{2,0}_E$ is an infinitesimal automorphism
of $\pi^{2,0}_E$, that is, $\mathcal{L}_{s_f}\pi^{2,0}_E=0$.
\end{proposition}

\begin{proposition}
If the almost complex Poisson structure $\pi^{2,0}_E$ on an almost complex
Lie algebroid $\left(E,\rho_{E},[\cdot,\cdot]_{E}, J_E\right)$ is
nondegenerated, then $J_E$ is integrable, and hence $E$ admits a $\partial_E$%
--symplectic structure.
\end{proposition}

\proof It is sufficient to prove that the bracket of two sections of type $%
(1,0)$ is again of type $(1,0)$; and this follows because, by Proposition %
\ref{p3.2}, the {non-degeneracy} condition implies that the mapping $\left(
\pi _{E}^{2,0}\right) ^{\#}$ is an isomorphism, so the space $\mathcal{V}%
^{1,0}(E)$ is (locally) spanned by the Hamiltonian sections, for which it is
satisfied. \qed

If $\pi_E^{2,0}$ is an almost complex {Poisson} structure on an almost
complex Lie algebroid $\left(E,\rho_{E},[\cdot,\cdot]_{E}, J_E\right)$ and $%
\pi_E=\pi_E^{2,0}+\overline{\pi_E^{2,0}}$ is the associated real Poisson
structure on $E$, then the distribution 
\begin{equation*}
\mathcal{D}=\cup_{x\in M}\mathcal{D}(x)\subset TM,\,\,\mathrm{where}\,\,%
\mathcal{D}(x):=\rho_E\left(\pi_E^\#(E_x^*)\right)\subset T_xM,\,x\in M
\end{equation*}
is a generalized foliation on $M$ (in the sense of Sussmann).

As well as we seen previously, for an almost complex Poisson structure $%
\pi^{2,0}_E$ on the almost complex Lie algebroid $\left(E,\rho_{E},[\cdot,%
\cdot]_{E}, J_E\right)$ such that $J_E$ is integrable, we have defined the
operator $\sigma_E^1:\mathcal{V}^{p,q}(E)\rightarrow\mathcal{V}%
^{p+1,q}(E)\oplus\mathcal{V}^{p+2,q-1}(E)$, by $\sigma_E^1(S)=-\left[%
S,\pi^{2,0}_E\right]_E$, for $S\in\mathcal{V}^{p,q}(E)$. Using the
properties of the Schouten-Nijenhuis bracket on $E$, it is easy to prove {%
that}:

\begin{enumerate}
\item[(i)] $\sigma_E^1\circ\sigma_E^1=0$,

\item[(ii)] $\sigma_E^1(S_1\wedge S_2)=\sigma_E^1(S_1)\wedge S_2+(-1)^{%
\mathrm{deg}(S_1)}S_1\wedge\sigma_E^1(S_2)$,

\item[(iii)] $\sigma _{E}^{1}\left( [S_{1},S_{2}]_{E}\right) =-[\sigma
_{E}^{1}(S_{1}),S_{2}]_{E}-(-1)^{\mathrm{deg}(S_{1})}[S_{1},\sigma
_{E}^{1}(S_{2})]_{E}$, for every $S_{1},S_{2}\in \mathcal{V}_{\mathbb{C}%
}^{\bullet }(E)$, where $\mathrm{deg}(S)$ denotes the degree of the
multisection $S$.
\end{enumerate}

Then, if we consider the decomposition of $\sigma _{E}^{1}$ induced by the
bigraduation, 
\begin{equation*}
\sigma _{E}^{1}=\sigma _{E}^{11}+\sigma _{E}^{12},
\end{equation*}%
and since $\sigma _{E}^{1}\circ \sigma _{E}^{1}=0$, it follows that $\sigma
_{E}^{11}\circ \sigma _{E}^{11}=\sigma _{E}^{12}\circ \sigma _{E}^{12}=0$
and $\sigma _{E}^{11}\circ \sigma _{E}^{12}+\sigma _{E}^{12}\circ \sigma
_{E}^{11}=0$. So, we obtain a differential bigraded complex 
\begin{equation*}
\ldots \overset{\sigma _{E}^{11}}{\longrightarrow }\mathcal{V}^{p-1,q}(E)%
\overset{\sigma _{E}^{11}}{\longrightarrow }\mathcal{V}^{p,q}(E)\overset{%
\sigma _{E}^{11}}{\longrightarrow }\mathcal{V}^{p+1,q}(E)\overset{\sigma
_{E}^{11}}{\longrightarrow }\ldots
\end{equation*}%
whose cohomology groups, denoted by $H_{CLP}^{p,q}(E)$, will be called 
\textit{complex Lichnerowicz-Poisson cohomology groups} of the almost
complex Poisson Lie algebroid $\left( E,\rho _{E},[\cdot ,\cdot
]_{E},J_{E},\pi _{E}^{2,0}\right) $ with $J_{E}$ integrable.

\subsection{Morphisms of almost complex Poisson Lie algebroids}

In this subsection, we consider the morphisms between almost complex Lie
algebroids preserving the almost complex Poisson structures.

\begin{definition}
Let $\left(E_1,\rho_{E_1},[\cdot,\cdot]_{E_1}, J_{E_1},
\pi^{2,0}_{E_1}\right)$ and $\left(E_2,\rho_{E_2},[\cdot,\cdot]_{E_2},
J_{E_2}, \pi^{2,0}_{E_2}\right)$ be two almost complex Poisson Lie
algebroids over the same base manifold $M$. A Lie algebroid morphism $%
\phi:E_1\rightarrow E_2$ is called an \textit{almost complex Poisson morphism%
} if:

\begin{enumerate}
\item[(i)] $\phi$ is an almost complex morphism;

\item[(ii)] $\phi$ is a Poisson type morphism with respect to $%
\pi_{E_1}^{2,0}$ and $\pi_{E_2}^{2,0}$, that is, $\phi$ satisfies one of the
following equivalent properties:

\begin{enumerate}
\item[1)] $\pi_{E_1}^{2,0}$ and $\pi_{E_2}^{2,0}$ are $\phi$--related, that
is 
\begin{equation*}
\pi_{E_1}^{2,0}(\phi^*(\varphi),\phi^*(\psi))=\pi_{E_2}^{2,0}(\varphi,\psi),%
\,\forall\,\varphi,\psi\in\Omega^{1,0}(E_2);
\end{equation*}

\item[2)] $\left(\pi_{E_2}^{2,0}\right)^\#=\phi\circ\left(\pi_{E_1}^{2,0}%
\right)^\#\circ\phi^*$.
\end{enumerate}
\end{enumerate}
\end{definition}

\begin{remark}
If $\phi:\left(E_1, J_{E_1}, \pi^{2,0}_{E_1}\right)\rightarrow\left(E_2,
J_{E_2}, \pi^{2,0}_{E_2}\right)$ is an almost complex Poisson morphism then $%
\phi:\left(E_1, \pi^{2,0}_{E_1}+\overline{\pi^{2,0}_{E_1}}%
\right)\rightarrow\left(E_2, \pi^{2,0}_{E_2}+\overline{\pi^{2,0}_{E_2}}%
\right)$ is an almost Poisson morphism of real Poisson Lie algebroids.
\end{remark}

Now, we can induce almost complex Poisson structures on Lie subalgebroids.

Let $\left(E,\rho_{E},[\cdot,\cdot]_{E}\right)$ be a Lie algebroid over a
smooth manifold $M$. According to \cite{Mack2}, a \textit{Lie subalgebroid}
over $M$, is given by a vector subbundle $p^\prime:E^{\prime}\rightarrow M$
such that:

\begin{enumerate}
\item[(i)] The anchor $\rho_E:E\rightarrow TM$ restricts to $%
p^\prime:E^\prime\rightarrow M$;

\item[(ii)] If $s^\prime_1,s^\prime_2\in\Gamma(E^\prime)$ then $%
[s^\prime_1,s^\prime_2]_E\in\Gamma(E^\prime)$ also.
\end{enumerate}

Let $\phi:\left(E_1,\rho_{E_1},[\cdot,\cdot]_{E_1}\right)\rightarrow%
\left(E_2,\rho_{E_2},[\cdot,\cdot]_{E_2}\right)$ be a $M$--morphism of Lie
algebroids over $M$. Then, according to \cite{I-M-D-M-P}, if $\phi$ is
injective then $\left(E_1,\rho_{E_1},[\cdot,\cdot]_{E_1}\right)$ is a 
\textit{Lie subalgebroid} of $\left(E_2,\rho_{E_2},[\cdot,\cdot]_{E_2}%
\right) $ over the same base $M$.

\begin{definition}
A Lie subalgebroid $(\widetilde{E},\phi)$ of an almost complex Poisson Lie
algebroid $\left(E, J_{E}, \pi^{2,0}_{E}\right)$, is said to be an \textit{%
almost complex Poisson Lie subalgebroid} of $E$ if the injective morphism $%
\phi:\widetilde{E}\rightarrow E$ is an almost complex Poisson morphism.
\end{definition}

\begin{definition}
A Lie subalgebroid $\widetilde{E}$ of an almost complex Poisson Lie
algebroid $\left(E, J_{E}, \pi^{2,0}_{E}\right)$ is said to be:

\begin{enumerate}
\item[(i)] $J_E$--\textit{coisotropic} if $J_E(\widetilde{E})\subset 
\widetilde{E}$ and 
\begin{equation}  \label{IV1}
\left(\pi_{E}^{2,0}\right)^\#\left(\mathrm{Ann}\,\widetilde{E}%
^{1,0}\right)\subset\widetilde{E}^{1,0},
\end{equation}
where $\mathrm{Ann}\,\widetilde{E}^{1,0}=\{\varphi\in(\widetilde{E}%
^{1,0})^*\,|\,\varphi(s)=0\,,\,\forall\,s\in\widetilde{E}^{1,0}\}$.

\item[(ii)] $J_E$--\textit{Lagrangian} if $J_E(\widetilde{E})\subset%
\widetilde{E}$ and 
\begin{equation}  \label{IV2}
\left(\pi_{E}^{2,0}\right)^\#\left(\mathrm{Ann}\,\widetilde{E}^{1,0}\right)=%
\widetilde{E}^{1,0}\cap\left(\pi_{E}^{2,0}\right)^\#\left((\widetilde{E}%
^{1,0})^*\right).
\end{equation}
\end{enumerate}
\end{definition}

As usual, we have

\begin{theorem}
A $M$--morphism $\phi:\left(E_1,\rho_{E_1},[\cdot,\cdot]_{E_1},J_{E_1},%
\pi_{E_1}^{2,0}\right)\rightarrow\left(E_2,\rho_{E_2},[\cdot,%
\cdot]_{E_2},J_{E_2},-\pi_{E_2}^{2,0}\right)$ between two almost complex
Poisson Lie algebroids is an almost complex Poisson morphism if and only if $%
\mathrm{Graph}\,\phi$ is a $J_E$--coisotropic Lie subalgebroid of $%
E=E_1\times E_2$, where $J_E$ is the product structure given in Example \ref%
{e3.4}.
\end{theorem}

\proof
We consider the almost complex Poisson Lie algebroids $\left(E_1,%
\rho_{E_1},[\cdot,\cdot]_{E_1},J_{E_1},\pi_{E_1}^{2,0}\right)$ and $%
\left(E_2,\rho_{E_2},[\cdot,\cdot]_{E_2},J_{E_2},-\pi_{E_2}^{2,0}\right)$
and the product structure described in Example \ref{e3.4} $(E=E_1\times E_2,$
$J_E=J_{E_1}+J_{E_2},\pi_{E}^{2,0}=\pi_{E_1}^{2,0}-\pi_{E_2}^{2,0})$.

Since $\mathrm{Graph}\,\phi=\{(s,\phi(s))\,|\,s\in\Gamma(E_1)\}$ is a
regular subalgebroid of $E=E_1\times E_2$, it is $J_E$--invariant if and
only if $(J_{E_1}(s),J_{E_2}(\phi(s)))\in \mathrm{Graph}\,\phi$, for any $%
s\in\Gamma(E_1)$, or equivalently, $(J_{E_2}\circ\phi)(s)=(\phi\circ
J_{E_1})(s)$, which means that $\phi$ is an almost complex morphism of Lie
algebroids.

Now, let us assume the previous statement, that is, $J_{E}$ be an almost
complex structure on $\mathrm{Graph}\,\phi $. Then, 
\begin{equation*}
\mathrm{Ann}\,(\mathrm{Graph}\,\phi )^{1,0}=\{(-\phi ^{\ast }(\varphi
),\varphi )\,|\,\varphi \in (E_{2}^{1,0})^{\ast }\},
\end{equation*}%
and by a straightforward calculation we have that $\mathrm{Graph}\,\phi $
satisfies \eqref{IV1}, and hence it is $J_{E}$--coisotropic, that is, 
\begin{equation*}
\left( (\pi _{E_{1}}^{2,0})^{\#}-(\pi _{E_{2}}^{2,0})^{\#}\right) \left( 
\mathrm{Ann}\,(\mathrm{Graph}\,\phi )^{1,0}\right) \subset (\mathrm{Graph}%
\,\phi )^{1,0},
\end{equation*}%
if and only if, 
\begin{equation*}
\left( \pi _{E_{2}}^{2,0}\right) ^{\#}(\varphi )=\left( \phi \circ \pi
_{E_{1}}^{2,0}\circ \phi ^{\ast }\right) (\varphi )\,,\,\forall \,\varphi
\in \Omega ^{1,0}(E_{2}),
\end{equation*}%
that is, $\phi $ is a Poisson type morphism. \qed

\section{Some other examples}

A \emph{general bisection} on $E$ is a $\mathcal{C}^{\infty }(M)$-valued $2$%
-cochain $F$ of $E^{\ast }$. It can be regarded as well as defined
equivalently by:

\begin{enumerate}
\item[--] a vector bundle map $F^{\#}:E^{\ast }\rightarrow E$, or

\item[--] a section $F$ of the vector bundle $E\otimes E$.
\end{enumerate}

The formula (\ref{I8}) can be used to define a non-necessary skew-symmetric
map $[\cdot ,\cdot ]_{F}:\Gamma \left( E^{\ast }\right) \times \Gamma \left(
E^{\ast }\right) \rightarrow \Gamma \left( E^{\ast }\right) $, 
\begin{equation}
\lbrack \omega ,\theta ]_{F}=\mathcal{L}_{F^{\#}\omega }\theta -\mathcal{L}%
_{F^{\#}\theta }\omega -d_{E}(F(\omega ,\theta )).  \label{I8a}
\end{equation}

Let $G$ be an endomorphism of $E$, $G^{\ast }:E^{\ast }\rightarrow E^{\ast }$
be the dual endomorphism of $G$ and $F$ be a general bisection. Then
consider the map $[\cdot ,\cdot ]_{F}:\Gamma \left( E^{\ast }\right) \times
\Gamma \left( E^{\ast }\right) \rightarrow \Gamma \left( E^{\ast }\right) $, 
\begin{equation}
\lbrack \omega ,\theta ]_{F}^{G}=\left[ G^{\ast }\omega ,\theta \right] _{F}+%
\left[ \omega ,G^{\ast }\theta \right] _{F}-G^{\ast }\left[ \omega ,\theta %
\right] _{F},  \label{18b}
\end{equation}%
{for every} $\omega ,\theta \in \Gamma (E^{\ast })$. If $F,G,\omega ,\theta $
are as above, denote $C\left( F,G)(\omega ,\theta \right) =[\omega ,\theta
]_{FG}-[\omega ,\theta ]_{F}^{G^{\ast }}$, where $FG$ is the general
bisection that corresponds to $\left( FG\right) ^{\#}=F\circ G^{\#}$. One
say that $F$ and $G$ are \emph{compatible} if $G\circ F^{\#}=F^{\#}\circ
G^{\ast }$ and $C(F,G)=0$.

If $G$ has a null Nijenhuis tensor $N_{G}=0$ and $F$ is a Poisson tensor,
then according to \cite{G-U, Kos, L-S-X} $(E,F,G)$ is a \emph{%
Poisson-Nijenhuis structure} if $F$ and $G$ are compatible.

If we suppose now that $J_{E}$ is integrable and $\pi _{E}$ is an almost
complex Poisson structure, then $J_{E}$ and $\pi _{E}$ are \emph{compatible}
if just $(E,\pi _{E},J_{E})$ is a Poisson-Nijenhuis structure. The
compatibility condition read

\begin{enumerate}
\item[1)] $J_{E}\circ \pi _{E}^{\#}=\pi _{E}^{\#}\circ J_{E}^{\ast }$;

\item[2)] $[\omega ,\theta ]_{\pi _{E}J}=[\omega ,\theta ]_{\pi
_{E}}^{J^{\ast }}$; {for every} $\omega ,\theta \in \Gamma \left( E^{\ast
}\right) $.
\end{enumerate}

We give below examples of integrable almost complex structures $J_{E}$ and
almost complex Poisson structures $\pi _{E}$ that are \textit{not compatible}
and that arise naturally on each sphere $S^{2n-1}$, $n\geq 1$. These
examples motivate the study of structures $J_{E}$ and $\pi _{E}$ that are
not necessary compatible. Notice that the compatibility of $J_{E}$ and $\pi
_{E}$ is not assumed in the results proved in our paper.

Let $n\geq 1$, $\bar{x}=\left( x^{\alpha }\right) _{\alpha =\overline{1,2n}%
}\in S^{2n-1}\subset I\!\!R^{2n}$, $\sum_{\alpha =1}^{2n}\left( x^{\alpha
}\right) ^{2}=1$, and $\bar{X}=\left( X^{\alpha }\right) _{\alpha =\overline{%
1,2n}}\in T_{\bar{x}}S^{2n-1}$, thus $\bar{X}\cdot \bar{x}=0$, i.e. $%
\sum_{\alpha =1}^{2n}x^{\alpha }X^{\alpha }=0$. We define the anchor $\rho
:S^{2n-1}\times I\!\!R^{2n}=TI\!\!R_{|S^{2n-1}}^{2n}\rightarrow TS^{2n-1}$
as the orthogonal projection $\rho _{\bar{x}}:T_{\bar{x}}I\!\!R^{2n}%
\rightarrow T_{\bar{x}}S^{2n-1}$, $(\forall )\bar{x}\in S^{2n-1}$, $\rho _{%
\bar{x}}(\bar{X})=\bar{X}-\left( \bar{X}\cdot \bar{x}\right) \bar{x}$.
Consider, for example, local coordinates for $\bar{x}\in S^{2n-1}$ as 
\begin{equation*}
\left( x^{\alpha }\right) _{\alpha =\overline{{1,2n}}}\rightarrow \left(
x^{1},\ldots ,x^{2n-1},t=\sqrt{1-\left( x^{1}\right) ^{2}-\cdots -\left(
x^{2n-1}\right) ^{2}}\right) .
\end{equation*}

Using these local coordinates,$\dfrac{\partial }{\partial x^{\alpha }}%
\rightarrow \bar{e}_{\alpha }-\dfrac{x^{\alpha }}{x^{2n}}\bar{e}_{2n}$. For $%
1\leq \alpha \leq 2n$ we have $\rho _{\bar{x}}(\bar{e}_{\alpha })=\dfrac{%
\partial }{\partial x^{\alpha }}-x^{\alpha }\bar{r}$ and $\rho _{\bar{x}}(%
\bar{e}_{2n})=-x^{2n}{\bar{r}}$, where $\bar{r}=\sum_{\alpha
=1}^{2n-1}x^{\alpha }\dfrac{\partial }{\partial x^{\alpha }}$.

For $1\leq \alpha <\beta \leq 2n-1$ we have $[\rho _{\bar{x}}(\bar{e}%
_{\alpha }),\rho _{\bar{x}}(\bar{e}_{\beta })]=-x^{\beta }\dfrac{\partial }{%
\partial x^{\alpha }}+x^{\alpha }\dfrac{\partial }{\partial x^{\beta }}$,
and $[\rho _{\bar{x}}(\bar{e}_{\alpha }),\rho _{\bar{x}}(\bar{e}_{2n})]=-t%
\dfrac{\partial }{\partial x^{\alpha }}$ for $\alpha =\overline{1,2n-1}$. It
follows that $\left( \mathcal{L}_{\bar{e}_{\alpha }}\bar{e}_{\beta }\right)
_{\bar{x}}=\left[ \bar{e}_{\alpha },\bar{e}_{\beta }\right] _{\bar{x}%
}=-x^{\beta }\bar{e}_{\alpha }+x^{\alpha }\bar{e}_{\beta }$, for $1\leq
\alpha <\beta \leq 2n$, extend to a bracket for sections of the vector
bundle $E=S^{2n-1}\times I\!\!R^{2n}\rightarrow S^{2n-1}$, according to the
anchor $\rho $.

It can be easily proved that $\left( E,\rho _{E},[\cdot ,\cdot ]_{E}\right) $
is a Lie algebroid.

Consider $n\geq 1$. The skew-symmetric matrix $J_{0}=\left( 
\begin{array}{cc}
0 & -I_{n} \\ 
I_{n} & 0%
\end{array}%
\right) $ is the same for three structures related to $E$: an integrable
complex structure $J$ on $E$, a Poisson bisection ${\widetilde{J}}:E^{\ast
}\rightarrow E$ and the adjoint $J^{\ast }:E^{\ast }\rightarrow E^{\ast }$
of $J$. The image of $\rho \circ {\widetilde{J}}$ gives a regular foliation
of $S^{2n-1}$ of codimension $1$. All these can be proved by straightforward
verifications.

For $n=1$, the foliation of the Lie algebroid is trivial (by points) on $%
S^{1}$, since $\rho \circ {\widetilde{J}}$ is null in this case.

For $n=2$, the Poisson structure induced by the Poisson bisection $\tilde{J}$
has the matrix%
\begin{equation*}
A\cdot J_{0}\cdot A^{t}=\left( 
\begin{array}{ccc}
0 & yz-tx & x^{2}+z^{2}-1 \\ 
tx-yz & 0 & tz+xy \\ 
-x^{2}-z^{2}+1 & -tz-xy & 0%
\end{array}%
\right) ,
\end{equation*}%
where 
\begin{equation*}
A=\left( 
\begin{array}{cccc}
1-x^{2} & -xy & -xz & -xt \\ 
-xy & 1-y^{2} & -yz & -yt \\ 
-xz & -yz & 1-z^{2} & -zt%
\end{array}%
\right)
\end{equation*}

The image of $\rho \circ {\widetilde{J}}$ gives a regular foliation of $%
S^{3} $ of codimension $1$. This remains true for any $n\geq 1$: $\rho \circ 
{\widetilde{J}}$ gives a regular foliation of $S^{2n-1}$ of codimension $1$.

The fact that $J$ is integrable follows by a straightforward computation of
the Nijenhuis tensor $N_{J}$ of $J$, that shows that $N_{J}=0$.

We study now the compatibility condition between $J$ and ${\widetilde{J}}$,
for $n\geq 1$. Consider $1\leq \alpha ,\beta ,\gamma \leq n$. Then: 
\begin{equation*}
\left( \mathcal{L}_{\bar{e}_{\alpha }}\bar{e}^{\beta }\right) \left( \bar{e}%
_{\gamma }\right) =-\bar{e}^{\beta }\left( \left[ \bar{e}_{\alpha },\bar{e}%
_{\gamma }\right] \right) =\delta _{\alpha }^{\beta }x^{\gamma }-\delta
_{\gamma }^{\beta }x^{\alpha }\,,\,\left( \mathcal{L}_{\bar{e}_{\alpha }}%
\bar{e}^{\beta }\right) \left( \bar{e}_{\gamma +n}\right) =-\bar{e}^{\beta
}\left( \left[ \bar{e}_{\alpha },\bar{e}_{\gamma +n}\right] \right) =\delta
_{\alpha }^{\beta }x^{\gamma +n},
\end{equation*}
which implies 
\begin{equation*}
\mathcal{L}_{\bar{e}_{\alpha }}\bar{e}^{\beta }=\delta _{\alpha }^{\beta
}\sum_{\gamma =1}^{n}\left( x^{\gamma }\bar{e}^{\gamma }+x^{\gamma +n}\bar{e}%
^{\gamma +n}\right) -x^{\alpha }\bar{e}^{\beta }.
\end{equation*}
Analogously, 
\begin{equation*}
\mathcal{L}_{\bar{e}_{\alpha }}\bar{e}^{\beta +n}=-x^{\alpha }\bar{e}^{\beta
+n}\,,\,\mathcal{L}_{\bar{e}_{\alpha +n}}\bar{e}^{\beta }=-x^{\alpha +n}\bar{%
e}^{\beta }\,,\,\mathcal{L}_{\bar{e}_{\alpha +n}}\bar{e}^{\beta +n}=\delta
_{\alpha }^{\beta }\sum_{\gamma =1}^{n}\left( x^{\gamma }\bar{e}^{\gamma
}+x^{\gamma +n}\bar{e}^{\gamma +n}\right) -x^{\alpha +n}\bar{e}^{\beta +n}.
\end{equation*}
We {also} have 
\begin{equation*}
\left[ \bar{e}^{\alpha },\bar{e}^{\beta }\right] _{{\widetilde{J}}}=\mathcal{%
L}_{{\widetilde{J}}^{\#}\bar{e}^{\alpha }}\bar{e}^{\beta }-\mathcal{L}_{{%
\widetilde{J}}^{\#}\bar{e}^{\beta }}\bar{e}^{\alpha }-d_{E}\left( {%
\widetilde{J}}\left( \bar{e}^{\alpha },\bar{e}^{\beta }\right) \right) =%
\mathcal{L}_{\bar{e}_{\alpha +n}}\bar{e}^{\beta }-\mathcal{L}_{\bar{e}%
_{n+\beta }}\bar{e}^{\alpha }=-x^{\alpha +n}\bar{e}^{\beta }+x^{\beta +n}%
\bar{e}^{\alpha }.
\end{equation*}
Analogously, 
\begin{eqnarray*}
&&\left[ \bar{e}^{\alpha },\bar{e}^{\beta +n}\right] _{{\widetilde{J}}%
}=2\delta _{\alpha }^{\beta }\sum_{\gamma =1}^{n}\left( x^{\gamma }\bar{e}%
^{\gamma }+x^{\gamma +n}\bar{e}^{\gamma +n}\right) -x^{\alpha +n}\bar{e}%
^{\beta +n}-x^{\beta }\bar{e}^{\alpha }, \\
&&\left[ \bar{e}^{\alpha +n},\bar{e}^{\beta }\right] _{{\widetilde{J}}%
}=-2\delta _{\alpha }^{\beta }\sum_{\gamma =1}^{n}\left( x^{\gamma }\bar{e}%
^{\gamma }+x^{\gamma +n}\bar{e}^{\gamma +n}\right) +x^{\alpha }\bar{e}%
^{\beta }+x^{\beta +n}\bar{e}^{\alpha +n}, \\
&&\left[ \bar{e}^{\alpha +n},\bar{e}^{\beta +n}\right] _{{\widetilde{J}}%
}=x^{\alpha }\bar{e}^{\beta +n}-x^{\beta }\bar{e}^{\alpha +n}.
\end{eqnarray*}
Using the formula 
\begin{equation*}
\left[ \omega ,\theta \right] _{{\widetilde{J}}}^{J^{\ast }}=\left[ J^{\ast
}\omega ,\theta \right] _{{\widetilde{J}}}+\left[ \omega ,J^{\ast }\theta %
\right] _{{\widetilde{J}}}-J^{\ast }\left[ \omega ,\theta \right] _{{%
\widetilde{J}}},
\end{equation*}
we have: 
\begin{eqnarray*}
\left[ \bar{e}^{\alpha },\bar{e}^{\beta +n}\right] _{{\widetilde{J}}%
}^{J^{\ast }}&=& \left[ J^{\ast }\bar{e}^{\alpha },\bar{e}^{\beta +n}\right]
_{{\widetilde{J}}}+ \left[ \bar{e}^{\alpha },J^{\ast }\bar{e}^{\beta +n}%
\right] _{{\widetilde{J}}}-J^{\ast }\left[ \bar{e}^{\alpha },\bar{e}^{\beta
+n}\right] _{{\widetilde{J}}} \\
&=&\left[ \bar{e}^{\alpha +n},\bar{e}^{\beta +n}\right] _{{\widetilde{J}}}-%
\left[ \bar{e}^{\alpha },\bar{e}^{\beta }\right] _{{\widetilde{J}}} \\
&&-J^{\ast }\left( 2\delta _{\alpha }^{\beta }\sum_{\gamma =1}^{n}\left(
x^{\gamma }\bar{e}^{\gamma }+x^{\gamma +n}\bar{e}^{\gamma +n}\right)
-x^{\alpha +n}\bar{e}^{\beta +n}-x^{\beta }\bar{e}^{\alpha }\right) \\
&=&x^{\alpha }\bar{e}^{\beta +n}-x^{\beta }\bar{e}^{\alpha +n}-\left(
-x^{\alpha +n}\bar{e}^{\beta }+x^{\beta +n}\bar{e}^{\alpha }\right) \\
&&-x^{\alpha +n}\bar{e}^{\beta }+x^{\beta }\bar{e}^{\alpha +n}-2\delta
_{\alpha }^{\beta }\sum_{\gamma =1}^{n}\left( x^{\gamma }\bar{e}^{\gamma
+n}-x^{\gamma +n}\bar{e}^{\gamma }\right) \\
& =&x^{\alpha }\bar{e}^{\beta +n}-x^{\beta +n}\bar{e}^{\alpha }-2\delta
_{\alpha }^{\beta }\sum_{\gamma =1}^{n}\left( x^{\gamma }\bar{e}^{\gamma
+n}-x^{\gamma +n}\bar{e}^{\gamma }\right).
\end{eqnarray*}

We have that $J{\widetilde{J}}:E^{\ast }\rightarrow E$ corresponds to the
matrix $J_{0}^{2}=-I_{2n}$, thus 
\begin{equation*}
\left( J{\widetilde{J}}\right) ^{\#}\left( \bar{e} ^{\alpha }\right) =-\bar{e%
}_{\alpha }\,\,\mathrm{and}\,\,\left( J{\widetilde{J}}\right) ^{\#}\left( 
\bar{e}^{\alpha +n}\right) =-\bar{e}_{\alpha +n}.
\end{equation*}
{It follows}: 
\begin{equation*}
\left[ \bar{e}^{\alpha },\bar{e}^{\beta +n}\right] _{J{\widetilde{J}}}=%
\mathcal{L}_{-\bar{e}_{\alpha }}\bar{e}^{\beta +n}-\mathcal{L}_{-\bar{e}%
_{\beta +n}}\bar{e}^{\alpha }=-\mathcal{L}_{\bar{e}_{\alpha }}\bar{e}^{\beta
+n}+\mathcal{L}_{\bar{e}_{\beta +n}}\bar{e}^{\alpha }=x^{\alpha }\bar{e}%
^{\beta +n}-x^{\beta +n}\bar{e}^{\alpha }.
\end{equation*}

For $n=1$, {we have} $\left[ \bar{e}^{1},\bar{e}^{2}\right] _{{\widetilde{J}}%
}^{J^{\ast }}=-x^{1}\bar{e}^{2}+x^{2}\bar{e}^{1}$ and $\left[ \bar{e}^{1},%
\bar{e}^{2}\right] _{J{\widetilde{J}}}=x^{1}\bar{e}^{2}-x^{2}\bar{e}^{1}$;
thus replacing ${\widetilde{J}}$ by $-{\widetilde{J}}$, the compatibility
condition can be fulfilled if $n=1$, but it is not the case if $n>1$.

\noindent Paul Popescu\newline
Department of Applied Mathematics\newline
University of Craiova\newline
Address: Craiova, 200585, Str. Al. Cuza, No. 13, Rom\^{a}nia\newline
email:\textit{paul$_{-}$p$_{-}$popescu@yahoo.com}

\end{document}